\documentclass[preprint,11pt]{aastex}
\pdfoutput=1
\usepackage{amsmath} \usepackage{amsfonts} \usepackage{amssymb}
\usepackage{apjfonts} \usepackage{color} \usepackage{graphicx}

\def\gsim{\;\rlap{\lower 2.5pt\hbox{$\sim$}}\raise 1.5pt\hbox{$>$}\;}
\def\lsim{\;\rlap{\lower 2.5pt\hbox{$\sim$}}\raise 1.5pt\hbox{$<$}\;}
\def\del{{\partial}}
\def\grad{\mbox{\boldmath{$\nabla$}}}
\def\bfOmega{\mbox{\boldmath{$\Omega$}}}

\def\bfzeta{\mbox{\boldmath{$\zeta$}}}
\def\beq{\begin{equation}}
\def\eeq{\end{equation}}
\newcommand{\HDblah}{HD~$\!$209458~$\!$b}

\shorttitle{EFFECTS OF INITIAL FLOW ON EXTRASOLAR PLANETS}
\shortauthors{THRASTARSON \& CHO}

\begin{document}

\title{Effects of Initial Flow on Close-In Planet Atmospheric
  Circulation}

\author{Heidar Th. Thrastarson and James Y-K.\ Cho\altaffilmark{1}}

\affil{Astronomy Unit, School of Mathematical Sciences, Queen Mary
  University of London, Mile End Road, London E1 4NS, UK}

\email{H.Thrastarson@qmul.ac.uk; J.Cho@qmul.ac.uk}

\altaffiltext{1}{Visiting scientist, Department of Terrestrial
  Magnetism, Carnegie Institution of Washington, Washington, DC 20015,
  USA}

\begin{abstract}

  We use a general circulation model to study the three-dimensional
  (3-D) flow and temperature distributions of atmospheres on
  tidally synchronized extrasolar planets.  In this work, we focus on
  the sensitivity of the evolution to the initial flow state, which
  has not received much attention in 3-D modeling studies.  We find
  that different initial states lead to markedly different
  distributions---even under the application of strong forcing (large
  day-night temperature difference with a short ``thermal drag time'')
  that may be representative of close-in planets.  This is in contrast
  with the results or assumptions of many published studies.  In
  general, coherent jets and vortices (and their associated
  temperature distributions) characterize the flow, and they evolve
  differently in time, depending on the initial condition.  If the
  coherent structures reach a quasi-stationary state, their spatial
  locations still vary.  The result underlines the fact that
  circulation models are currently unsuitable for making quantitative
  predictions (e.g., location and size of a ``hot spot'') without
  better constrained, and well posed, initial conditions.
 
\end{abstract}


\section{Introduction}

Understanding the flow dynamics of atmospheres is crucial for
characterizing extrasolar planets. Dynamics strongly influence the
temperature distribution as well as the spectral behavior.  An
essential tool for studying dynamics on the large-scale is a global
hydrodynamics model.  Many studies have used such a model
\citep[e.g.,][]{Showman02,Choetal03,Cooper05, Langton07,
  Choetal08,Dobbs-Dixon08,Showmanetal08a,Menou09}.  The models in
these studies numerically solve a set of non-linear partial
differential equations for the evolution of a fluid on a rotating
sphere.  Hence, the initial condition (as well as the boundary
conditions) needs to be specified.

Presently, physically accurate and mathematically well-posed initial
conditions for the models are not known for extrasolar planets. Unlike
for the solar system planets, dynamically ``balanced'' initial data\footnote{
self-consistent set of fields which does not lead to
excessive noise and deviations from accurate prediction} are not
available and dominant dynamical processes, such as baroclinic
instability and geostrophic turbulence, are not yet understood for the
extrasolar planets \citep{Choetal03,Choetal08, Showmanetal08b,Cho08}.
Concerning initialization, there is a long history of research in
geophysical fluid dynamics and numerical weather prediction, and it is
still a subject of active research---even for the Earth
\citep{Holton04}.

In most simulations of close-in planets performed so far, the initial
state is either at rest or with a small, randomly perturbed wind field
to break the flow symmetry \citep{Cooper05,Langton07,
  Dobbs-Dixon08,Showmanetal08a, Menou09}. \citet{Choetal08} initialize
their two-dimensional simulations with random eddies, and variations
of the initial velocity distributions are studied. They find
significant differences in the flow evolution, depending on the vigor
of the eddies. On the other hand, \citet{Cooper05} report on a
three-dimensional (3-D) simulation, set up with an initial retrograde
equatorial jet, and find no qualitative difference, compared with one
starting from a rest state. \citet{Showmanetal08b} and \citet{Cho08}
give summaries of the various results.

In this work, we present runs from an advanced 3-D general circulation
model. As in \citet{Cooper05}, as well as in \citet{Showmanetal08a}
and \citet{Menou09}, the model used in this work solves the full
primitive equations \citep{Pedlosky87}. However, there are some
important assets in the model used in this work (see section \ref{sec:method}), 
compared with most models used so far. 
For example, it uses a parallel pseudospectral algorithm
\citep{Orszag70,Eliasenetal70, Canutoetal88} with better-controlled,
less invasive numerical viscosity. In this regard, our model is
similar to the one used by \citet{Menou09}.

With our model, we focus on the sensitivity of the flow evolution to
the initial state. The sensitivity has not been much emphasized in
previous studies, particularly in those using 3-D circulation models. In
order to unambiguously delineate the sensitivity effect, we set up the
simulations in a manner similar to previous studies, apply idealized
forcing (in many cases unencumbered by a vertical variation), and
compare runs with all parameters identical---except for the initial
condition.

The basic plan of the paper is as follows. We describe the model and
its setup for our simulations in section \ref{sec:method}, where we endeavor
to provide enough details to facilitate reproduction of the results.
In section \ref{sec:results} we present the results of simulations
initialized with different organized large-scale flow patterns,
including the rest state. In this section, we also show how sensitive
the flow is to small perturbations in the initial wind field. We
conclude in section \ref{sec:concl}, summarizing this work and discussing
its implications for close-in extrasolar planet circulation modeling
work.

\section{Method}\label{sec:method}

\subsection{Governing Equations}\label{subsec:eqns}

The global dynamics of a shallow, 3-D atmospheric layer is governed by
the primitive equations \cite[e.g.,][]{Pedlosky87,Holton04}. Here, by
``shallow'' we mean the thickness of the atmosphere under
consideration is small compared to the planetary radius $R_p$. In
atmospheric studies, pressure $p$ is commonly used as the vertical
coordinate. In the pressure coordinate system, these equations read:
\begin{subequations}\label{eq:pe} \begin{eqnarray}
  \frac{{\rm D}{\bf v}}{{\rm D}t} + 
  \left(\frac{u}{R_p}\tan\phi\right){\bf k}\times{\bf v}\ & = &\
  -\grad\! _p \Phi - f{\bf k}\times{\bf v} + {\cal D}_{\bf v}\\
  \frac{\del\Phi}{\del p}\ & = &\ -\frac{1}{\rho}\\
  \frac{\del\omega}{\del p}\ & = &\ -\grad\! _p\cdot{\bf v}\\ 
  \frac{{\rm D}T}{{\rm D} t} -\frac{\omega}{\rho\, c_p}\ & = &\ 
  \frac{\dot{q}_{\rm{\tiny net}}}{c_p} + {\cal D}_T\, ,
\end{eqnarray}
\end{subequations}
where
\begin{eqnarray*}
  \frac{{\rm D}}{{\rm D}t}\ =\ \frac{\del}{\del t} + 
  {\bf v}\cdot\grad\! _p + \omega\!\frac{\del}{\del p}\, .
\end{eqnarray*}

In the above equations, ${\bf v}({\bf x},t)\! =\! (u,v)$ is the
(eastward, northward) velocity in a frame rotating with $\Omega$, the
planetary rotation rate; $\Phi = gz$ is the geopotential, where $g$ is
the gravitational acceleration and $z$ is the distance above the
planetary radius $R_p$; ${\bf k}$ is the unit vector in the local
vertical direction; $f = 2 \Omega\sin\phi$ is the Coriolis parameter,
the projection of the planetary vorticity vector $2\bfOmega$ onto
${\bf k}$, with $\phi$ the latitude; $\grad\!  _p$ is the horizontal
gradient on a constant $p$-surface; $\omega\equiv{\rm D}p/{\rm D}t$ is
the vertical velocity; $\rho$ is the density; ${\cal D}_{\bf v} = -
\nu\grad^4{\bf v}$ represents the momentum dissipation, with $\nu$ the
constant viscosity coefficient; $T$ is the temperature; $c_p$ is the
specific heat at constant pressure; $\dot{q}_{\rm{\tiny net}}$ is the
net diabatic heating rate; and, ${\cal D}_T = - \nu\grad^4T$
represents the temperature dissipation.

Equations~(\ref{eq:pe}) are closed with the ideal gas law, $p = \rho
RT$, as the equation of state, with $R$ the specific gas constant.  A
suitable set of boundary conditions, used in this work, is ${\rm
  D}p/{\rm D}t = 0$ at the top and bottom $p$-surfaces.  Hence, the
boundaries are material surfaces and no mass flow is allowed to cross
the boundaries.  With these boundary conditions, the equations admit
the full range of motions for a stably-stratified atmosphere---except
for sound waves.  For a discussion of the various aspects of the
primitive equations (including superviscosity) and their use for
extrasolar planet application, the reader is referred to
\citet{Choetal03}, \citet{Choetal08} and Cho \& Gulsen (in
preparation).  In this work, as described below,
equations~(\ref{eq:pe}) are actually solved in a more general
coordinate system\footnote{which is useful when variations in the
  bottom boundary, caused by static or dynamic conditions, are not
  small}.

\subsection{Numerical Model}\label{subsec:numerical}

To solve equations~(\ref{eq:pe}) in the spherical geometry, we use the
Community Atmosphere Model (CAM 3.0). CAM is a well-tested,
highly-accurate pseudospectral hydrodynamics model developed by the
National Center for Atmospheric Research (NCAR) for the atmospheric
research community \citep{Collinsetal04}. For hydrodynamics problems
not involving sharp discontinuities (e.g., shocks) and irregular
geometry, the pseudospectral method is superior to the standard grid
and particle methods \citep[e.g.,][]{Canutoetal88}.

As in many pseudospectral formulations of the algorithm, CAM solves
the equations in the vorticity-divergence form in the horizontal
direction, where $\bfzeta = \grad\times{\bf v}$ is the vorticity and
$\delta = \grad\cdot{\bf v}$ is the divergence. In the vertical
direction, CAM uses the generalized $p$-coordinate:
\begin{equation}\label{eq:eta}
  p(\lambda,\phi,\eta,t)\ =\ A(\eta) p_r\, +\, 
  B(\eta) p_s(\lambda,\phi,t)\, ,  
\end{equation}
where $\lambda$ is the longitude, $\phi$ is the latitude, $\eta$ is
the generalized vertical coordinate, $p_r$ is a constant reference
pressure, $p_s(\lambda,\phi,t)$ is a deformable pressure surface at
the bottom boundary, and $A,B\in[0,1]$. In the vertical direction, CAM
uses the finite difference method. Superviscosity ($\nabla^4$
operators), as well as a small Robert-Asselin time filter $\epsilon$
\citep{Robert66,Asselin72}, are applied at every timestep in each
layer to stabilize the integration. The timestepping is done using a
semi-implicit, second-order leapfrog scheme. Note that effects of
various numerical dissipation are often subtle and can be significant
on the integration, particularly over long times
\citep[e.g.,][]{Dritscheletal07}. Further details of the model and the
effects of numerical viscosity on the flow evolution will be described
elsewhere.

\subsection{Model Setup}

In all the simulations discussed in this paper, the physical
parameters chosen are based on the close-in extrasolar planet,
HD209458b.  The basic result presented---that the evolution depends on
the initial flow state---does not change for a different close-in
planet.  The physical parameters for the model HD209458b planet are
listed in Table~\ref{tab:params}.

CAM is able to include radiatively-active species and their coupling
to the dynamics.  However, we do not include them in the present work
so that the effects discussed are not obfuscated by complications
unrelated to the essential result.  Our principle motivation is to
study the dependence on the initial flow in the most unambiguous way
possible.  To this end, the flow is forced using the simple Newtonian
drag formalism, as in many previous studies of extrasolar planet
atmospheres \citep[e.g.,][]{Cooper05,Langton07,
  Showmanetal08a,Menou09}.  This drag is a simple representation of
the net heating term in equation~(\ref{eq:pe}d):
\begin{eqnarray}
  \frac{\dot{q}_{\rm{\tiny net}}}{c_p}\  =\ 
  -\frac{1}{\tau_{\rm th}}\left(T - T_e\right),
\end{eqnarray} 
where $T_e = T_e(\lambda,\phi,\eta, t)$ is the ``equilibrium''
temperature distribution and $\tau_{\rm th}$ is the thermal drag time
constant.  

In this work, both $T_e$ and $\tau_{\rm th}$ are prescribed and
barotropic ($\del/\del\eta = 0$) and steady ($\del/\del t = 0$),
although simulations relaxing these restrictions have been run to
verify robustness of our results. In general, both $T_e$ and
$\tau_{\rm th}$ are (as are $R$ and $c_p$) complicated functions of
space and time \citep{Cho08}. Here, 
\begin{equation}\label{eq:Te}
  T_e\ =\ T_m + \Delta T_e \cos\phi \cos\lambda\, ,
\end{equation}
where $T_m = (T_D + T_N)/2$ and $\Delta T_e = (T_D - T_N)/2$ and $T_D$
and $T_N$ are the maximum and minimum temperatures at the day and
night sides, respectively. Most of the simulations described in this
paper have $T_D$ = 1900 K, $T_N$ = 900 K, and $\tau_{\rm th} =
3$~\HDblah\ planet days (where $\tau_p \equiv 2\pi/\Omega$ is 1~planet
day). Note that we have varied the timescale of the forcing by using a
$\tau_{\rm th}$ value in the range from 0.01 to 10~planet days, as
well as letting the timescale to decrease with height. The main result
does not change for values $\gsim 0.1$ day, which nearly covers the entire
spectrum of $\tau_{\rm th}$ in all past studies using the Newtonian
drag formalism.

The spectral resolution in the horizontal direction for most of the
runs described in the paper is T42, which corresponds to 128$\times$64
grid points in physical space\footnote{Note that, because of the
higher order accuracy of the spectral method, this essentially corresponds 
to a finite difference resolution of over 420$\times$210, for smooth fields.}.  We have
performed runs with resolutions varying from T21 (64$\times$32) to T85
(256$\times$128), in order to check convergence of the solutions.  The
vertical direction is resolved by 26 coupled layers, with the top
level of the model located at 3~mbar.\footnote{Table~\ref{tab:eta} in
  the Appendix gives the positions of all the model levels (layer
  interfaces).} The pressure at the bottom $\eta$ boundary is
initially 1~bar, but the value of the pressure changes in time. This
range of pressure is chosen because it encompasses the region where
current observations are likely to be probing and where most of the
circulation modeling studies have thus far directed their attention.
We have also performed simulations in which the domain extends down to
100~bars and again verified that the basic behavior described in this
paper is not affected. The entire domain is initialized with an
isothermal temperature distribution, $T_m$ = 1400~K.

\section{Results}\label{sec:results}

\subsection{Basic Dependence: Jets}\label{sec:basic}

\begin{figure} 
  \centerline{\plotone{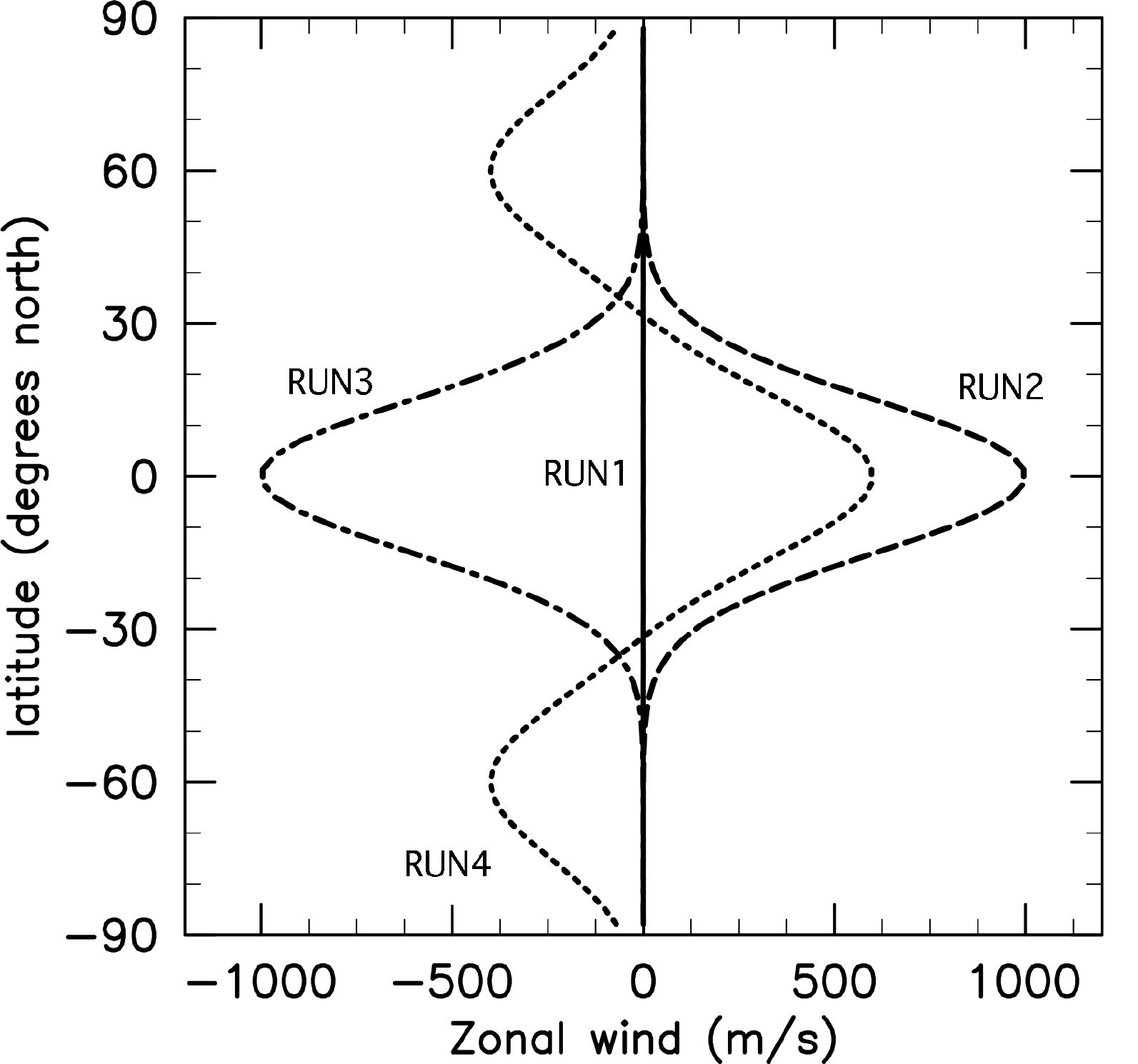}} 
  \caption{Initial conditions for simulations RUN1 (---), RUN2
    (-\,-\,-), RUN3~($-\,\cdot\,-$), and RUN4 ($\cdots$).  The
    height-independent, zonally-symmetric, eastward velocities,
    $u(\lambda,\phi,\eta,0) = u_0(\phi)$, are
    shown.  \label{fig:Uini}}
\end{figure}

To examine the robustness of evolved flow states to organized initial
flow configurations, we have performed simulations with a wide range
of initial conditions.  The conditions from four of those runs
(labeled RUN1--RUN4) are shown in Figure~\ref{fig:Uini}.  In all the
runs presented, the setup is identical---except for the initial flow
configuration.  The physical and numerical parameters/conditions are
given in Tables~\ref{tab:params} and \ref{tab:sims}, respectively.

RUN1 is initialized with a small, random perturbation introduced in
the flow. Specifically, values of $u$ and $v$ are drawn from a
Gaussian random distribution centered on zero with a standard
deviation of 0.05~m~s$^{-1}$. RUN2 is initialized with a
zonally-symmetric, eastward equatorial jet of the following form:
\begin{equation} 
  \label{eq:uini} 
  u_0(\phi)\ =\ U 
  \exp\left\{\frac{(\phi - \phi_0)^2}{2 \sigma^2}\right\},
\end{equation}
where $u(t\! =\! 0)\, =\, u_0$, $U = 1000$~m~s$^{-1}$, $\phi_0 = 0$,
and $\sigma = \pi / 12$. RUN3 is initialized with a westward
equatorial jet described by equation~(\ref{eq:uini}), with $U =
-1000$~m~s$^{-1}$, $\phi_0 = 0$, and $\sigma = \pi / 12$. RUN4 is
initialized with a flow containing three jets. Note that the condition
for RUN4 is very similar to the zonal average of the wind field of
RUN1 at 50~planetary rotations. The jet profiles presented in
Figure~\ref{fig:Uini} are independent of height, as well as longitude.

\begin{figure} 
  \plotone{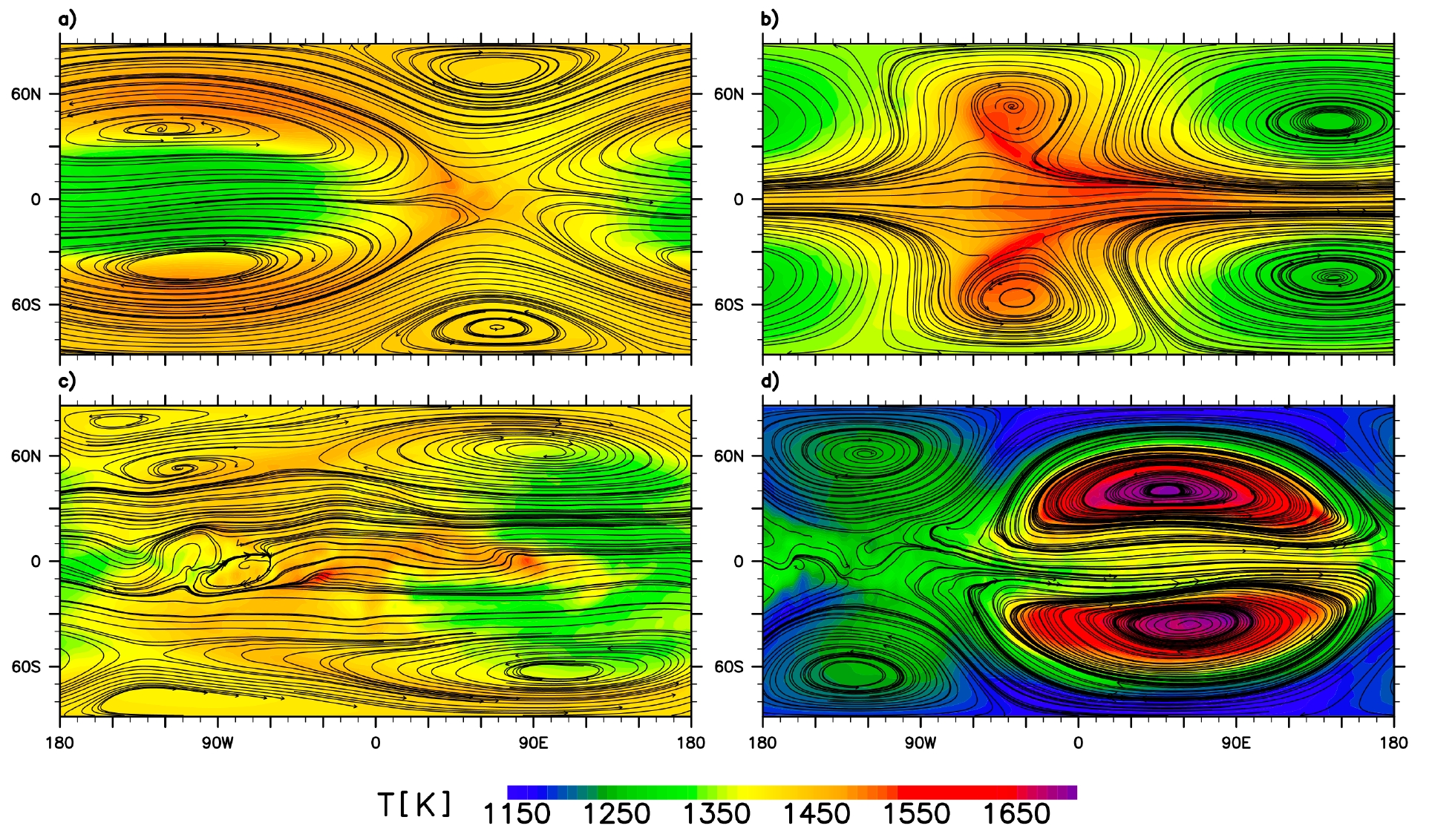} 
  \caption{Temperature (color map) and flow (streamlines) fields at 
    $t= 40\,\tau_p$, near the $p \sim 900$~mbar level, for RUN1 (a), 
    RUN2 (b), RUN3 (c), and RUN4 (d).  The fields are shown in
    cylindrical-equidistant projection centered at the equator.  The 
    four simulations are set up identically, except for the initial 
    wind field.  The location and size of vortices, and the associated    
    temperature patterns, strongly depend on the initial wind 
    configuration.  
    \label{fig:Tlev22}}\footnote{bla}
\end{figure}

Figure~\ref{fig:Tlev22} shows the temperature and flow\footnote{ 
Here, and in other figures, streamlines are shown.  Streamlines are 
obtained by smoothly following the flow; 
they are tangent to the instantaneous velocity vectors at each grid point.} 
fields of the four runs at $t/\tau_p = 40$ (or $t/\tau_{\rm th} \approx\! 14$). 
The fields near the 900~mbar pressure level are shown. [Recall that the
$\eta$ level-surfaces of our model are functions of pressure, as
described by equation~(\ref{eq:eta}).] The figure illustrates the
major point of this paper: given different initial states, there are
clear, qualitative (as well as quantitative) differences between the
different runs. Qualitatively, there are some common features. For
example, most of the runs exhibit a coherent quadrupole flow
structure---two large cyclonic and anti-cyclonic vortex-pairs
straddling the equator.\footnote{The cyclonicity of a vortex is
  defined by the sign of $\bfzeta\cdot\bfOmega$: it is positive for a
  cyclone and negative for an anticyclone.} However, the {\it
  location} of an individual vortex is different in the runs---as is
the temperature pattern. In RUN3, a distinct quadrupole pattern is not
present but there are more vortices in this run compared to the other
runs. The temperature distributions are different because they are
strongly linked to the flow. Consequently, the minimum-to-maximum
temperature ranges vary from a moderately large 550~K (RUN4) to only
about 200~K (RUN3) in the figure.

\begin{figure} 
  \plotone{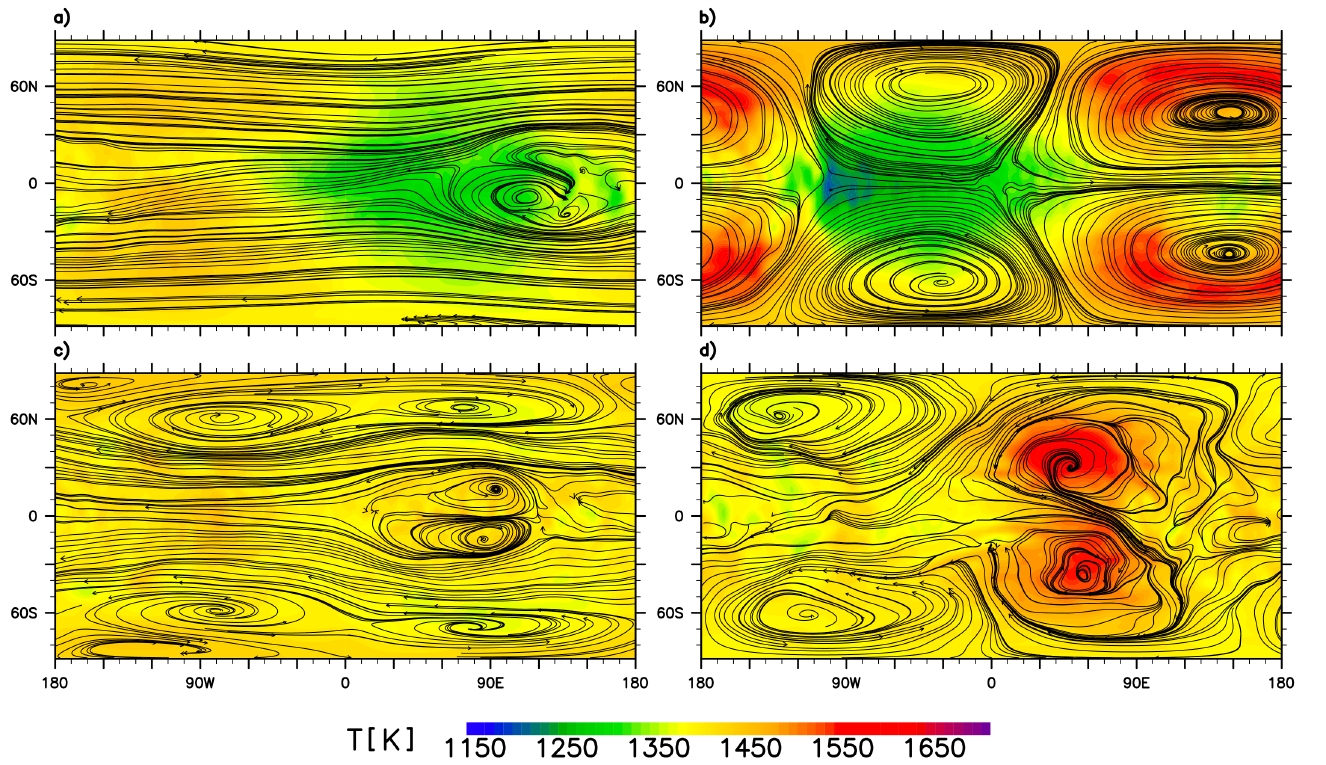}
  \caption{Temperature (color map) and flow (streamlines) fields at $t
    = 40\,\tau_p$, near the $p \approx$ 85~mbar level,
    for the same four simulations as shown in Figure~\ref{fig:Tlev22}: 
    RUN1 (a), RUN2 (b), RUN3 (c) and RUN4 (d).
    The sensitivity to the initial wind state is present throughout all
    heights in the atmosphere.
    \label{fig:Tlev7}}
\end{figure}

The behavior just described is not restricted to a single altitude.
Figure~\ref{fig:Tlev7} shows the fields corresponding to those
presented in Figure~\ref{fig:Tlev22}, but at a higher altitude ($p
\approx 85$~mbar pressure level). Comparison of
Figures~\ref{fig:Tlev22} and \ref{fig:Tlev7} illustrates the
structural differences in 3-D (vertical), as well as in 2-D
(horizontal). In RUN2 and RUN4, the large-scale vortices are strongly
aligned, forming columns through most of the height extent of the
modeled atmosphere; that is, the flow is strongly barotropic. In the
other two runs, the flow is not vertically aligned throughout in large
parts of the modeled atmosphere---and, therefore, the flow is
baroclinic. The two figures also point to the corresponding strong
difference in 3-D temperature distributions, associated with the flow
structures.

\begin{figure}
  \plotone{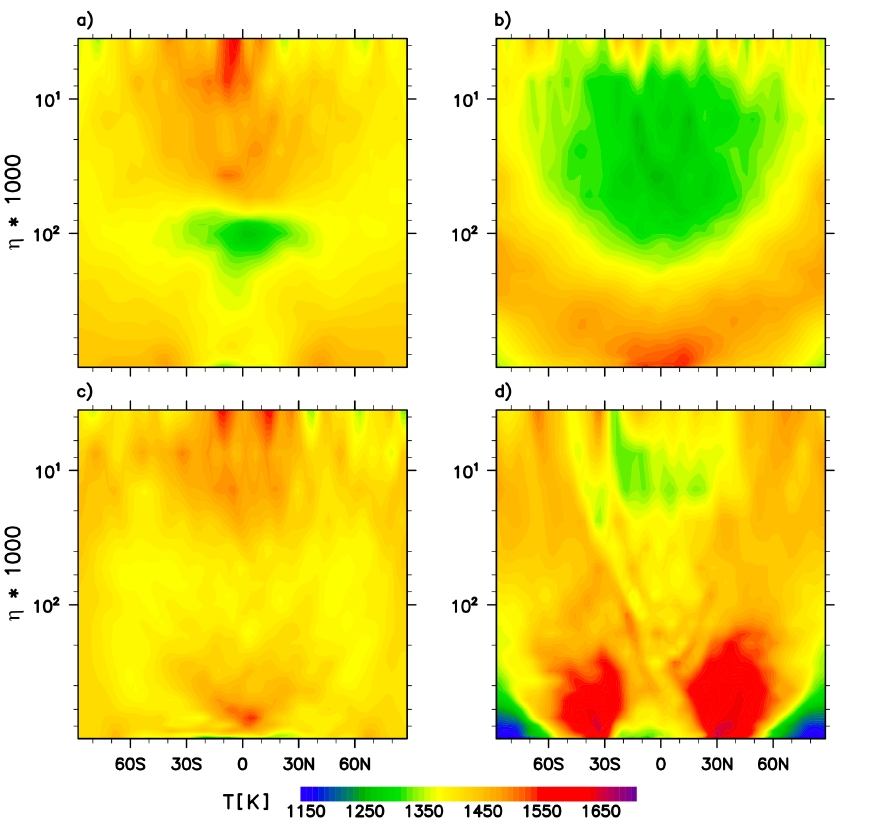}
  \caption{Snapshots of temperature vertical cross-section
    at the sub-stellar longitude, at $t = 40\,\tau_p$,
    for the four simulations presented in Figure~\ref{fig:Tlev22}:  
    RUN1 (a), RUN2 (b), RUN3 (c) and RUN4 (d).  
    The vertical temperature structure is strongly sensitive to the 
    initial flow state, and usually contains an inversion. 
    \label{fig:Tvert}}
\end{figure}

This is more clearly seen in Figure~\ref{fig:Tvert}. The figure shows
the vertical (height-latitude) cross-section of the temperature at 0
degrees (sub-stellar) longitude from the runs presented in Figures
\ref{fig:Tlev22} and \ref{fig:Tlev7}. In Figure~\ref{fig:Tvert}, the
hottest and coldest regions are at different locations in all the
runs. Near the equator, RUN1 exhibits a strong temperature
inversion\footnote{See \citet{Burrowsetal07} and \citet{Knutsonetal08}
  for discussion of thermal inversion in the context of close-in
  extrasolar giant planets.}, while RUN2 and RUN4 do not. In addition,
RUN2 and RUN4 exhibit generally strong decreases in temperature with
height, while the others do not. As can be seen, the degree of
temperature mixing varies strongly in the vertical direction among the
runs---from $\sim\! 200$~K contrast (RUN3) to $\sim\! 500$~K contrast
(RUN4). In general, the vertical structure is of low order, containing
usually a single inversion. In our study, some form of inversion
appears to be a generic feature.

Preliminary steady state analysis of the primitive equations suggests
that the basic behavior described above is due to the way in which the
applied, $(s,n) = (1,1)$, forcing projects onto the normal modes of
the planetary atmosphere; here, $s$ is the zonal wavenumber and $n$ is
the total (sectoral) wavenumber of the spectral harmonics. In
particular, a normal mode decomposition of the atmosphere into the
vertical structure and Hough functions
\citep[e.g.,][]{Chapman70,Longuet-Higgins67} indicates that the
forcing projects mostly onto low-order baroclinic modes, when the
initial state is at rest. In contrast, when the initial state contains
large-scale jets, the forcing projects more strongly on the barotropic
mode, compared to the runs started from rest. Similar behavior has
been observed in studies of the Earth's troposphere under tropical
forcing \citep[e.g.,][]{Geisler82,Lim83}. A more detailed study of
coupling between forcing and normal modes is currently being performed
and will be described elsewhere.

\begin{figure}
  \plotone{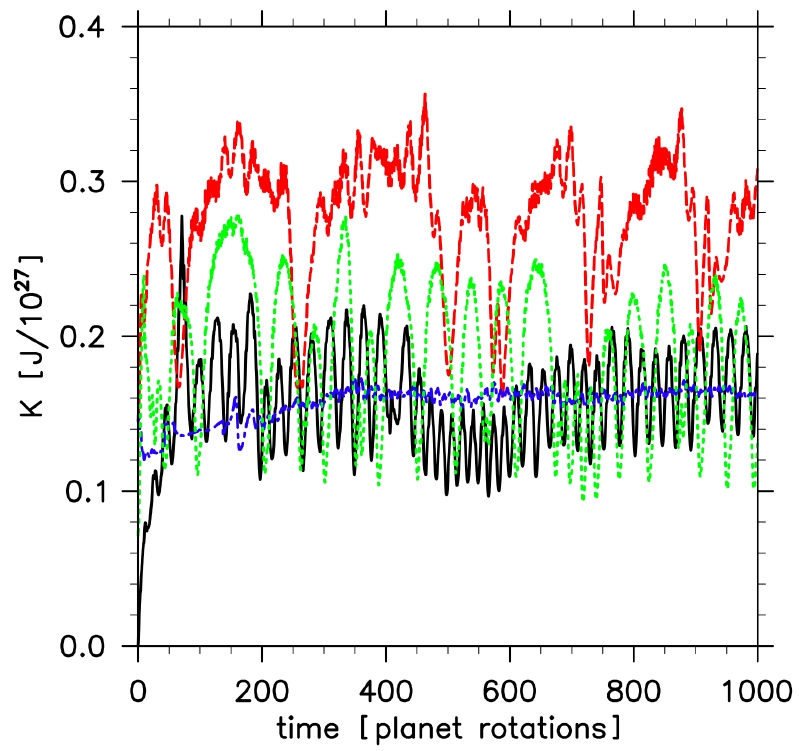}
  \caption{Time series of total kinetic energy, integrated over the 
     domain, for RUN1 (black ---), RUN2 (red --\,--\,--), 
     RUN3 (blue $-\,\cdot\,-$), and RUN4 (green -\,-\,-).
    \label{fig:kine}}
\end{figure}

Furthermore, it is important to note that all of the above features,
both dynamical and thermal, can vary in time. 
All of these features are important for observations 
\citep[e.g.,][]{Knutsonetal07} and spectral modeling 
\citep[e.g.,][]{Tinettietal07}.
A thorough study of the long-time evolution (over 1000~planetary rotations, 
or more than $330~\tau_{\rm{th}}$) of the runs reveals a fundamental 
difference in their temporal behavior as well. For example, the flow pattern 
in RUN2 is characterized by two vertically aligned vortex columns in each 
hemisphere that translate longitudinally around the poles. The temperature 
in the upper altitude region is more strongly coupled to the flow than it is in 
the lower altitude regions. The flow pattern in RUN4 is also a set of vertically
aligned vortex columns, but the columns oscillate in the east-west
direction. The patterns in RUN1 and RUN3 are more complex, exhibiting
a mixture of vortex splitting and merger and stationary states at
different altitude levels. 
Figure~\ref{fig:kine}, which shows a time series of the total kinetic 
energy, gives a quantitative measure of the different temporal behavior 
of the simulations.

\begin{figure}
  \plotone{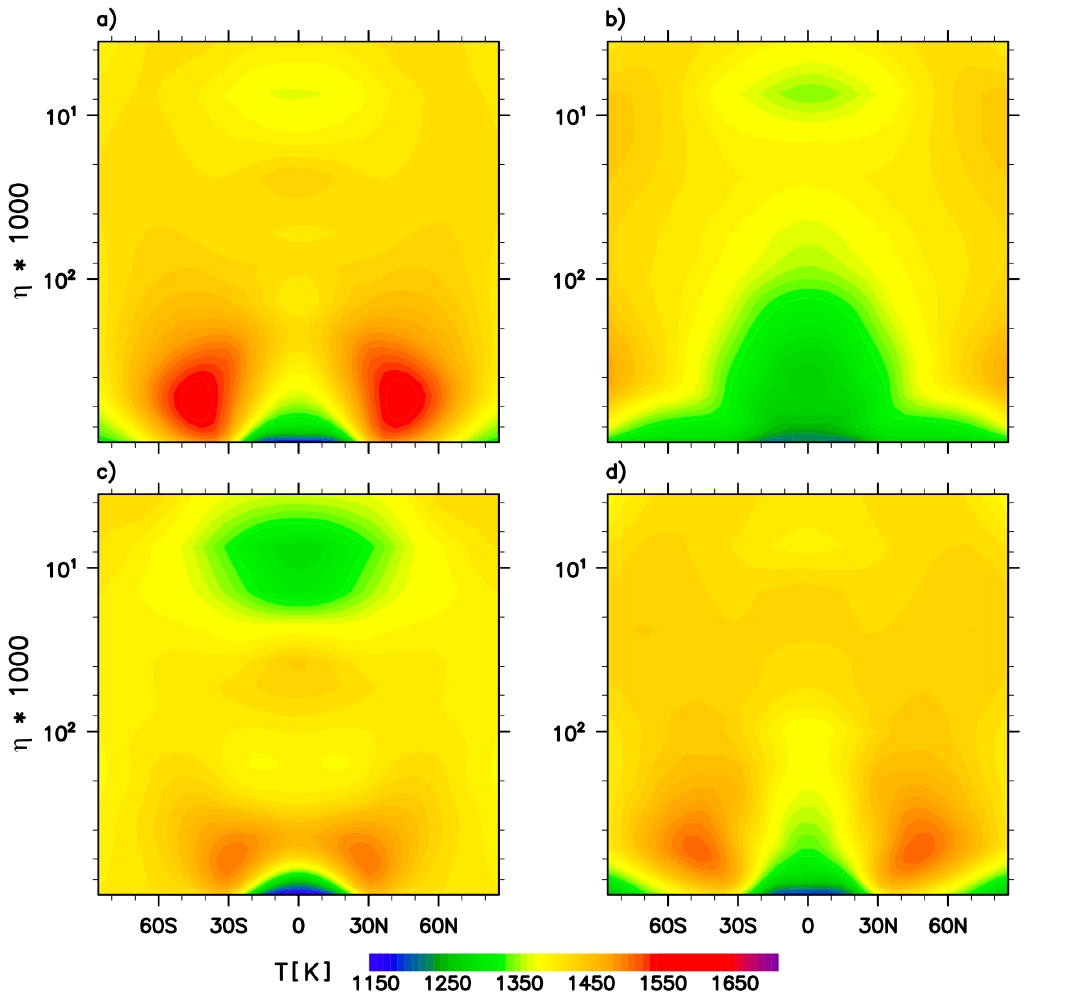}
  \caption{Temperature vertical cross-section at an arbitrary
    longitude ($\lambda = 135^{\circ}$), averaged over 450 planetary
    rotations (150 thermal drag times), for the four simulations presented in 
    Figure~\ref{fig:Tlev22}: RUN1 (a), RUN2 (b), RUN3 (c) and RUN4 (d).
    The difference in the temperature structure is independent of long
    time-averaging and is not due to a phase shift in a quasi-periodic 
    evolution.   
    \label{fig:T-tmean}}
\end{figure}

When the flow field is time-averaged
  over a long period, the time-mean state is also
  significantly sensitive to the initial flow. This can be seen in
  Figure~\ref{fig:T-tmean}, which shows temperature
  cross-sections at an arbitrary longitude ($\lambda = 135^\circ$),
  averaged over 450 planetary rotations (planet day 300 to 750).  
  The figure clearly shows that the variability observed is not simply 
  a result of a phase shift in a quasi-periodic evolution. The flow and
  temperature structures in each run are fundamentally different from
  one another.

Interestingly, the full range of flow and temperature behavior
described above has been previously captured qualitatively, using the
one-layer equivalent-barotropic model \citep{Choetal08}. The
equivalent-barotropic equations are a reduced, vertically-integrated
version of the full primitive equations used in this study
\citep{Salby89}. In many situations, the reduced model can be
fruitfully used to study the dynamics of the full model by varying the
Rossby deformation radius to represent different heights (or
temperatures) of the full multi-layer model
\citep[e.g.,][]{Choetal08,Scott08}, and this also appears
  to be the case for hot extrasolar planets.

\subsection{Extreme Sensitivity: Small Stirring}
\label{sec:sensitivity}

\begin{figure}
  \plotone{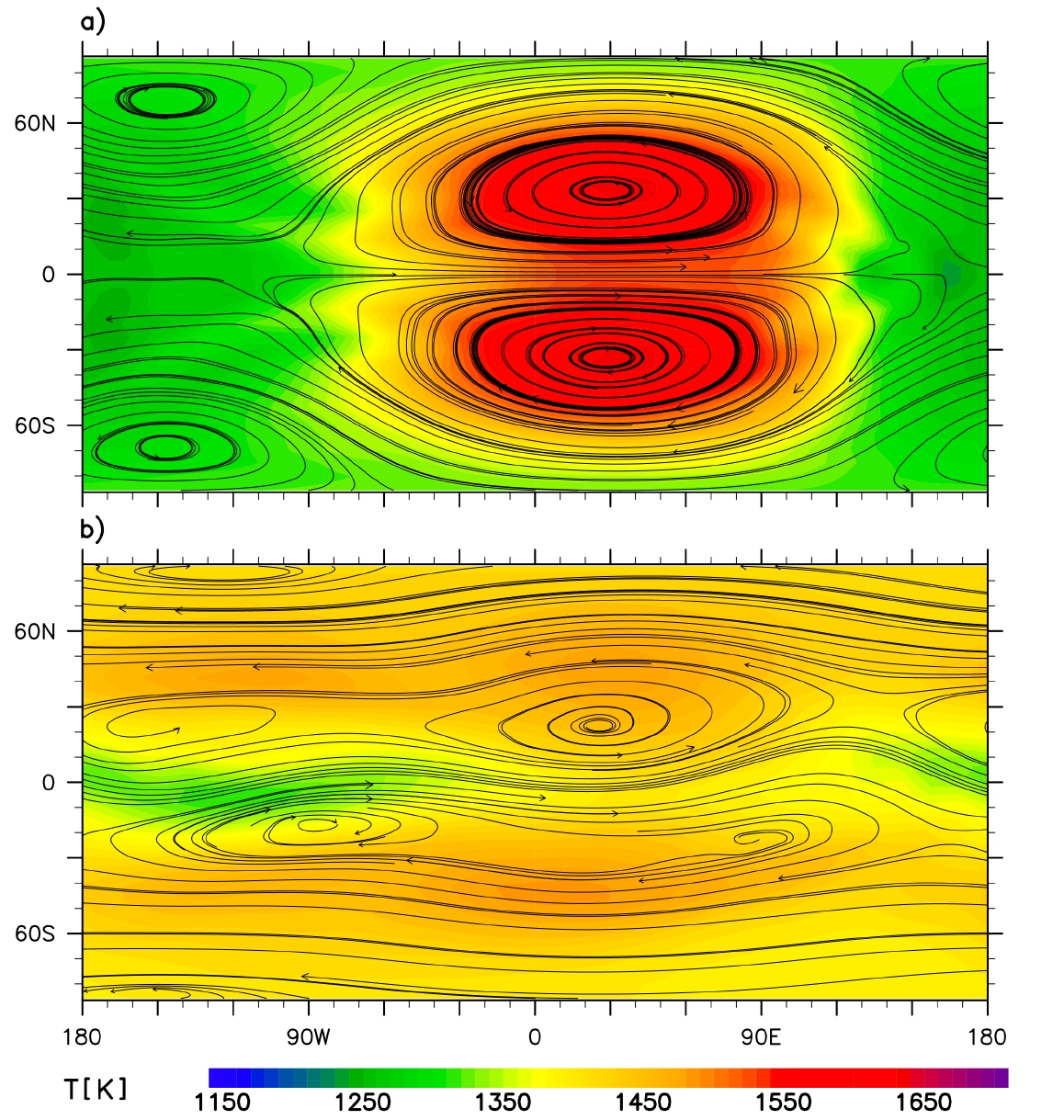}
  \caption{Temperature (color map) and flow (streamlines) fields 
    after 1000 planetary rotations, at the $p \sim 420$~mbar level, 
    for two simulations differing only by small deviations in the 
    initial wind.  
    The top panel shows the result of a simulation started from rest, 
    while the bottom panel is from a simulation started with a small 
    perturbation.  
    Note the clear asymmetry in the north-south direction, which only
    appears when the small perturbations are present in the initial state. 
    \label{fig:stir}}
\end{figure}

As might be expected from general nonlinear dynamics theory, in fact
the evolution can be strongly sensitive to small differences in the
initial flow state. This is illustrated in Figure~\ref{fig:stir}.
There, two simulations are presented (panels a and b), which are
identical in all respects except for a minute difference in the
initial flow. The simulation in the top panel (RUN5) is started from
rest. In contrast, the simulation in the bottom panel (RUN1) is
started with a small perturbation: the initial values of $u$ and $v$
at each grid point are set to a Gaussian random distribution, centered
on zero with a standard deviation of 0.05~m~s$^{-1}$.  Note that the
maximum initial wind perturbation magnitude is only about 0.02\% of
the typical root mean square flow speed in the frames shown
($\sim$ 500~m~s$^{-1}$).

The two panels in Figure~\ref{fig:stir} show temperature and flow
distributions after $t/\tau_p = 1000$ (or $t/\tau_{\rm th} \approx 333$)
at the $p \sim 420$~mbar level. 
The distributions in the two panels are clearly different.  
At times they may look more similar than shown here, but in general that is not the case. 
The two runs generally show a different temporal behavior. 
Note that in Figure~\ref{fig:stir}a, there is a high degree of hemispheric symmetry, 
particularly in the north-south direction. In contrast, Figure~\ref{fig:stir}b 
shows a clear asymmetry in both the north-south and east-west direction. 
The small asymmetry in Figure~\ref{fig:stir}a is entirely due to machine 
precision and is not physical, since there is no way to break the symmetry 
in the setup of the run. Therefore, not surprisingly, some mechanism for 
inducing a noticeable symmetry breaking is necessary. The salient point 
here is, however, that even a tiny perturbation can lead to a marked 
difference in the flow and temperature distributions, even at relatively 
early integration times.

\subsection{Robustness: Additional Parameter Variations}
\label{subsec:robustness}

\begin{figure}
  \plotone{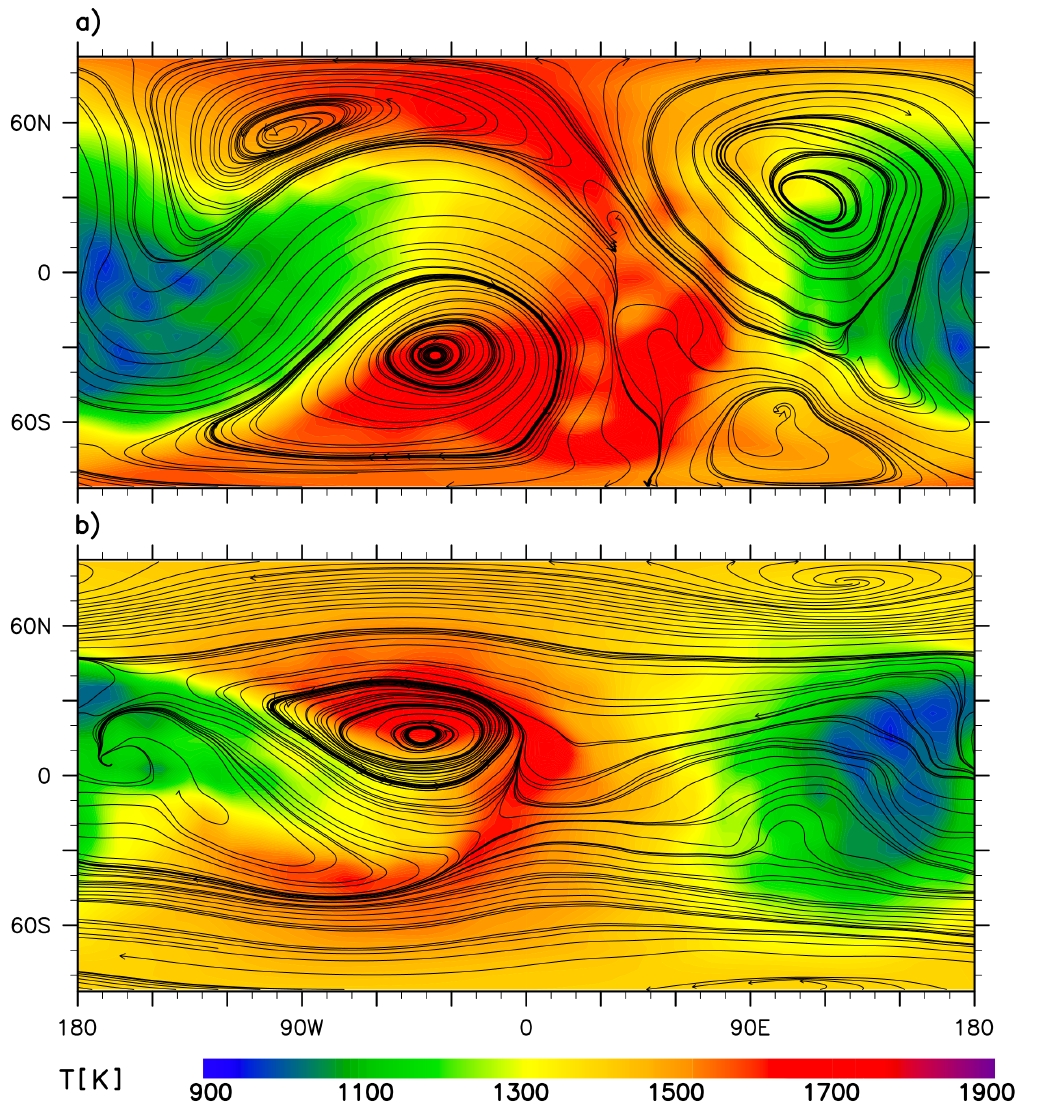}
  \caption{Snapshots of temperature fields (color coded) with
    streamlines overlaid.  Fields at the $p \sim$ 900 mbar  level 
    are shown at t = 1000 planetary rotations, for two simulations (RUN6 and
    RUN7); the thermal drag time is 0.5 planetary rotations (42 hours) in these runs.  
    The only difference between the simulations is the initial wind state.  
    The top panel shows the result of a simulation started with only a 
    small perturbation and the bottom panel is from a simulation
    started with a westward jet.  The sensitivity to the initial flow 
    state is still present for this small value of the 
    thermal drag time (cf., RUNS 1--4).
  \label{fig:tau42h}}
\end{figure}

It is important to understand that the dependence on the initial flow
state is robust and the behavior is not limited only to the
parameters, and the ranges, discussed thus far. The dependence has
been verified for numerous model parameters and ranges. For example,
Figure~\ref{fig:tau42h} shows that the strong dependence on the
initial wind exists for much shorter $\tau_{\rm th}$, despite the
strong forcing such drag times entail. In this case, $\tau_{\rm
  th}/\tau_p = 0.5$. The upper panel is from RUN6 and the lower panel
is from RUN7. The former run is initialized with only small stirring
and no organized jet. In contrast, the latter run is initialized with
a westward equatorial jet, identical to the setup of RUN3. At the
shown time and height, there are clear differences between the flow
and temperature patterns of the two simulations. The coldest area is
advected east of the anti-stellar point in RUN6, but west of the
anti-stellar point in RUN7. Furthermore, all the vortices have
different locations. In RUN7, there is a fairly zonally symmetric jet
at high latitudes, leading to a much more homogenized temperature
distribution above the mid-latitudes than in RUN6.

\begin{figure}
  \plotone{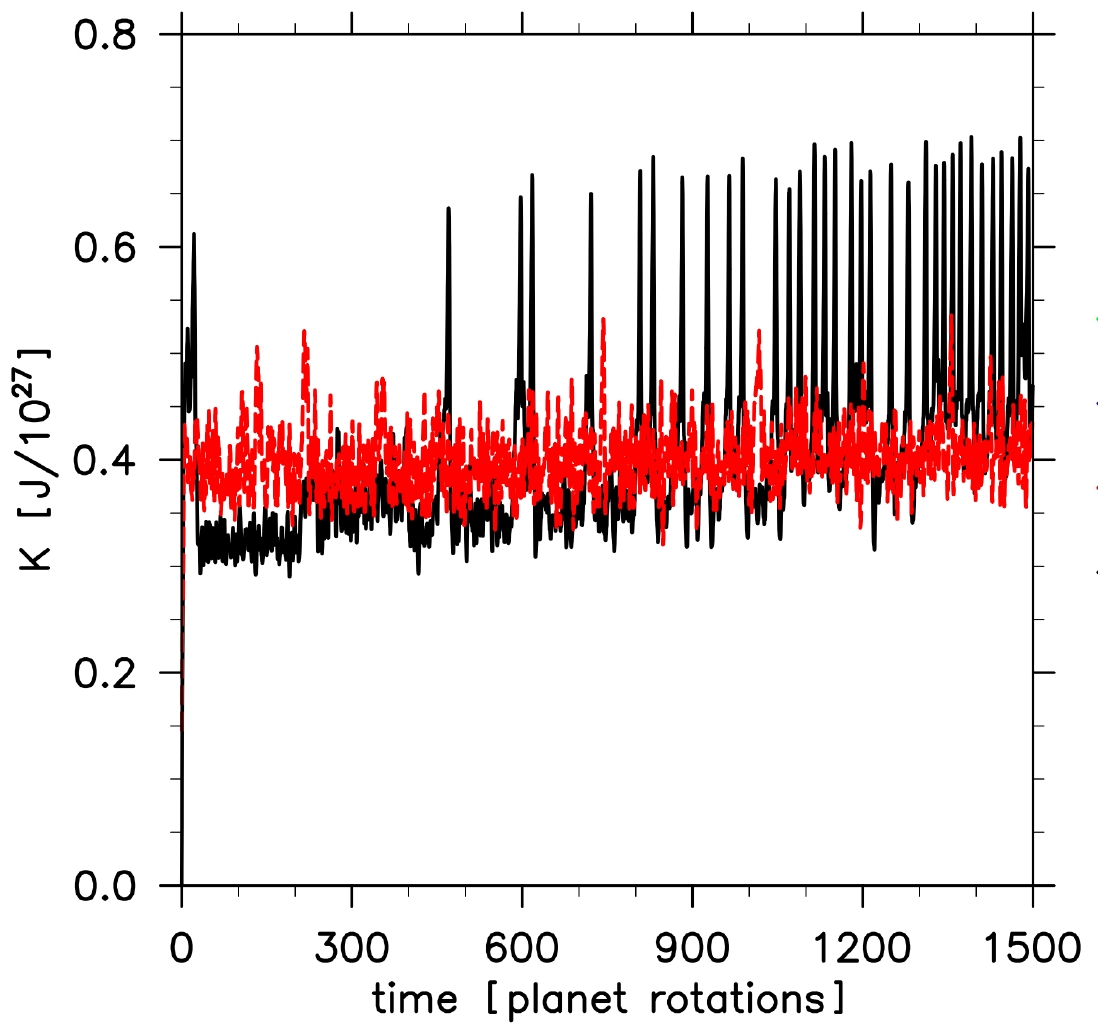}
  \caption{Time series of total kinetic energy, integrated over the domain, 
     for RUN6 (black ---) and RUN7 (red --\,--\,--).
  \label{fig:kine-tau42h}}
\end{figure}

While a weaker difference might be expected in this case, 
  based on other studies (e.g., Cooper \& Showman 2005), 
  the sensitivity is unabated in our simulations. And, this holds for the long 
  time average behavior as well. For a time mean over 300 rotations (e.g., 
  planet days 1200 to 1500), at the level shown in Figure~\ref{fig:tau42h}, 
  the location of the coldest region differs by 40 degrees in longitude 
  between the two simulations shown in the figure. 
  Figure~\ref{fig:kine-tau42h} shows a time series of the total kinetic 
  energy in the two simulations, revealing their different evolution. 
  Even after integrating for a very long time (15,000 rotations), we have 
  checked that the differences between RUN6 and RUN7 are still present. 
  Note that this is a much longer integration time than that reported in any 
  published studies of close-in planet circulation thus far. 
  However, the flow at such long times is inexorably affected by cumulative 
  numerical dissipation and phase errors and the result obtained should 
  not be taken too literally \citep[e.g.,][]{Canutoetal88}.

In addition, we have studied the dependence when the initial jet
contains a vertical shear--i.e., the jet distribution is baroclinic.
Simulations have been initialized with an eastward jet, which has zero
magnitude at the bottom and increases linearly with height so that
the lateral flow distribution in the top layer of the model is
identical to that in RUN2.\footnote{Recall that the flow distribution
  in this run is barotropic.} In the baroclinic case, vortex columns
evolve that extend throughout most of the atmosphere, as in RUN2.
However, unlike in RUN2, the temperature structure here is also
strongly barotropic. This appears to be related to the way the forcing
projects onto the free modes of the system, as described in
section \ref{sec:basic}.

As alluded to in section \ref{subsec:eqns}, the timescale of the fastest
motions admitted by equations~(\ref{eq:pe}) with the given boundary
condition is $\sim\! 40$~minutes, for the planetary physical parameters used in
this work. Hence, a forcing with timescales of the same order or
smaller is not entirely physically self-consistent with the use of
equations~(\ref{eq:pe}). Notably, such forcing introduces
numerical difficulties. 
For the physically unrealistic drag time of
$\tau_{\rm th}$ = 1~hour, for which excessive numerical dissipation is
required to prevent the code from blowing up or being inundated with
numerical noise, the {\it temperature} evolution is only mildly sensitive 
to the initial condition, at long-time integrations. 
However, when such short drag times are only applied over a limited 
range in pressure---i.e., $\tau_{\rm th}$ is allowed to vary with $\eta$---
the sensitivity to the initial condition is present even after long time 
integrations (7000 planet days).

\section{Conclusion}\label{sec:concl}

In this paper, we have shown that in a generic general circulation
simulation of a tidally synchronized planet, the flow and temperature
distributions depend strongly on the initial state. In all simulations
initialized with a different jet configuration, large scale coherent
vortices are formed; but, their location, size, and number varies,
depending on the initial wind. The temperature distribution is
relatively homogenized by the flow, compared to the large temperature
contrast in the forcing equilibrium temperature profile. But the
degree of mixing, as well as locations of temperature extremes vary
between differently initialized simulations. The time variability of
the atmosphere---i.e. how vortices and associated temperature patterns
move around the planet or whether they stay at fixed
positions---varies depending on the initial wind.

The Newtonian drag scheme used in this study is idealized and the
``correct'' parameters to use in the set-up are unknown, with many
choices possible. Explorations of different parts of the parameter
space will be presented elsewhere. Here, we have chosen the set-up to
capture, as cleanly as possible, the effects of the initial flow in a
regime of parameter space plausibly relevant for hot Jupiters with
strong zonally-asymmetric forcing. We have found that the strong
dependence on the initial wind is valid for a wide range of thermal
drag times ($\tau_{\rm th}$ = 0.5--10 planet days) and with T21, T42,
and T85 resolutions. Some reduction in the sensitivity is sometimes observed for
the very small $\tau_{\rm th}$ and long time integration simulations,
which require large numerical dissipation to prevent the fields from
being dominated by small scale noise. This situation is, however, 
not physically realistic and numerically suspect.

The strong dependence on initial wind has implications for the use of
general circulation models for interpretation of observations of
extrasolar planet atmospheres. These results underline that, while
numerical circulation models of the kind employed here are useful for
studying plausible mechanisms and flow regimes, they are currently
unsuitable for making ``hard'' predictions---such as exact locations
of temperature extremes on a given planet.

\acknowledgments 

H.\TH.\TH.\ is supported by the EU Fellowship. J.Y-K.C. is supported
by the NASA NNG04GN82G and STFC PP/E001858/1 grants. The authors thank
O. M. Umurhan and Chris Watkins for helpful discussions, and the anonymous 
referee for helpful suggestions.

\newpage

\appendix

\section{Appendix}

Table~\ref{tab:eta} presents the $A$ and $B$ coefficients of
equation~(\ref{eq:eta}) in \S\ref{subsec:numerical}. Note that, as
defined, each $\eta$-surface can span across a range of $p$-surfaces.

\clearpage

\begin{deluxetable}{llll}
  \tablecaption{{Physical Parameters}
    \label{tab:params}} 
  \tablehead{}
  \startdata
  Planetary rotation rate & $\Omega$ &  2.1$\times$10$^{-5}$  & s$^{-1}$ \\
  Planetary radius & $R_p$ &  10$^8$  & m \\ 
  Gravity & $g$  &  10  & m s$^{-2}$ \\ 
  Specific heat at constant pressure & $c_p$ & 1.23$\times$10$^4$ &
  J kg$^{-1}$ K$^{-1}$ \\
  Specific gas constant & $R$ &  3.5$\times$10$^3$ & J kg$^{-1}$ K$^{-1}$ \\
  \\
  Mean equilibrium temperature & $T_m$ &  1400 & K \\
  Equilibrium substellar temperature & $T_D$ &  1900 & K \\
  Equilibrium antistellar temperature & $T_N$ &  900 & K \\
  Initial temperature & $T_0$ & 1400 & K \\
 \enddata
\end{deluxetable}

\begin{deluxetable}{clccc} 
  \tablecaption{{Summary of Runs Discussed}
    \label{tab:sims}} 
  \tablehead{ RUN & Initial Flow & 
    $\tau_{\rm th}$ [$2\pi/\Omega$] & $N$ & 
    $\nu$\ [10$^{21}$~m$^4$~s$^{-1}$] } 
   \startdata
  1 & small noise  &  3    &  42  & 1  \\
  2 & eastward jet &  3    &  42  & 1  \\
  3 & westward jet &  3    &  42  & 1  \\
  4 & three jets   &  3    &  42  & 1  \\
  5 & zero winds   &  3    &  42  & 1  \\
  6 & small noise  &  0.5  &  21  & 10 \\
  7 & westward jet &  0.5  &  21  & 10 \\  
  \enddata
  \tablecomments{$\tau_{\rm th}$ is the thermal drag
    timescale, $N$ is the spectral truncation wavenumber, and $\nu$ is
    the superviscosity coefficient.  All the runs have Robert-Asselin
    filter coefficient $\epsilon$ = 0.06.  In the T42 runs the
    timestep is $\Delta t$ = 120~s, but for T21 resolution $\Delta t$
    = 240~s.}
\end{deluxetable}

\clearpage

\placefigure{f1}

\clearpage

 \begin{deluxetable}{lrr}
  \tablecaption{{Vertical Coordinate Coefficients\tablenotemark{a}}
     \label{tab:eta}} 
   \tablehead{ \colhead{Level Surface Index} &
     \colhead{A$\times 10^3$} & \colhead{B$\times10^3$} } 
   \startdata
  0 & 2.2 &  0 \\
   1 & 4.9 &  0 \\
   2 & 9.9 &  0 \\
   3 & 18.1 &  0 \\
   4 & 29.8 &  0 \\
   5 & 44.6 &  0 \\
   6 & 61.6 &  0 \\
   7 & 78.5 &  0 \\
   8 & 77.3 &  15.0 \\
   9 & 75.9 &  32.8 \\
   10 & 74.2 &  53.6 \\
   11 & 72.3 &  78.1 \\
   12 & 70.0 &  106.9 \\
   13 & 67.3 &  140.9 \\
   14 & 64.1 &  180.8 \\
   15 & 60.4 &  227.7 \\
   16 & 56.0 &  283.0 \\
   17 & 50.8 &  347.9 \\
   18 & 44.7 &  424.4 \\
   19 & 37.5 &  514.3 \\
   20 & 29.1 &  620.1 \\
   21 & 20.8 &  723.5 \\
   22 & 13.3 &  817.7 \\
   23 & 7.1 &  896.2 \\
   24 & 2.5 &  953.5 \\
   25 & 0 &  985.1 \\
   26 & 0 &  1000.0\\
   \enddata
    \tablenotetext{a}{The pressure at each point is given by
     equation~(\ref{eq:pe}) in the text, with $p_r$ = 1~bar.}
\end{deluxetable} 

\end{document}